\documentstyle[aps,pre,multicol,epsf]{revtex}

\begin{document}

\title{Correlations of triggering noise in driven magnetic clusters}

\author{Bosiljka Tadi\'c$^\star $}

\address{Jo\v{z}ef Stefan Institute,
P.O. Box 3000, 1001 Ljubljana, Slovenia }


\maketitle
\begin{abstract}
We show that the temporal fluctuations $\Delta H(t)$ of the
threshold driving field $H(t)$, which triggers an avalanche in slowly
driven disordered ferromagnets with many domains, exhibit long-range
correlations in space and time.
The probability distribution of the distance
between {\it successive} avalanches as well as the
distribution of trapping times of domain wall at a given point in space
have fractal properties  with the universal scaling exponents.
We show how these   correlations are
related to  the scaling behavior of Barkhausen avalanches occurring by
magnetization reversal. We also suggest a transport equation which
takes into account the observed noise correlations.
 \end{abstract}

\begin{multicols}{2}
\section{Introduction}
Field driven disordered ferromagnets at low temperatures exhibit
Barkhausen (BH) noise,
a very important physical phenomenon, which is used for noninvasive
characterization technique in commercial alloys.
It has been recognized that measured BH noise exhibits scale free
behavior when only the external field is varied for all values of
disorder and driving fields in real experiments
 \cite{MC1,TOR,E4,BGD} (a short summary of the experimental data  can be
found in Ref.\ \cite{BT-Calcutta}). This is in sharp contrast to some
theoretical conclusions \cite{cornell} emphasizing fine tuning of
the strength of disorder to a single critical value.
  It  has been also attempted to
understand the dynamics of domain walls which results in BH avalanches
in terms of the dynamics of a sandpile model \cite{GRB,Cuba} (for a recent
review of sandpile models see \cite{DD}) and of
the models of interface depinning \cite{E4,Brasil,Durin-stress,TN}.
It should be stressed that in BH noise the presence of disorder plays
an important role via pinning of domain walls. It remains to be
understood how the motion of domain walls is affected by the
pinning.

Different  scaling behaviors of BH noise can be attributed
to the domain structure, which is related to annealing and type of
impurities, and nucleation and coalescence of domains, as well sa
varying driving conditions.
Annealing the samples in the applied anisotropic stress or in
the magnetic field leads to a characteristic structure
of extended system-size domains with a $180^\circ$ domain walls
parallel to the anisotropy axis  \cite{LM,Brasil,Durin-stress}.
Numerical simulations with extended domain wall in two
dimensions  \cite{TN} led to
the conclusion that different scaling behavior can be expected
in two limiting  cases: (a) When disorder is weak BH response is
dominated by motion of a single (extended) domain wall; (b)
For strong disorder a multidomain structure occurs  with many
competing domain walls. This picture is in a qualitative agreement
with  experimental results in stress annealed
Fe-B-Si \cite{Brasil} and Fe-Co-B \cite{Durin-stress} alloys. Therefore
the two different universality classes can be related to surface and
bulk criticality, respectively.
Numerical simulations starting from an uniform ground state
\cite{cornell,Vives,BT}, on the other hand,
do not take into account extended domain walls, and thus correspond
to  the behavior at high-disorder. In both cases, however, the origin
of scaling has not been fully understood.

Recently the exact results of the random-field Ising model on the Bethe
lattice \cite{SD} show
that the  avalanche distributions at a fixed driving field have
finite cutoffs. However, an infinite cutoff appears for a range of
disorder strengths $\Delta <\Delta _c$ if a
distribution is integrated over the hysteresis loop. The integration
thus involves an infinite jump in magnetization (when system size
$L\to \infty$) at a critical value of driving field.
On the other hand, for strong
disorder $\Delta $ above $\Delta _c$ the avalanche distributions remain
exponentially damped for all fields \cite{SD}. These results
encourage further study of the field-integrated distributions to
elucidate the role of driving field in the appearance of the  scale-free
BH noise.

In this work we study properties of the threshold driving field
which triggers Barkhausen
avalanches in the multidomain structure, which is  generated in the
linear  part of the ascending branch of  hysteresis.
When the system is slowly driven by increasing the external magnetic
field $H$ with time, even by continuous field changes, there
exists a {\it threshold} field that equals weakest pinning force
of a domain wall in the system and thus starts an avalanche. In the
numerical experiment we can adjust
the driving field updates to the weakest local field in a system
(so-called infinitely slow driving), and thus we can examine  precisely
the properties of the triggering noise.
Surprisingly, we find that time series of such  field updates are
long-range correlated. Both Fourier spectrum of the threshold
field fluctuations $\Delta H(t) \equiv H(t+1)-H(t)$ and distribution of
distances of the successive
avalanches which that field triggers appear to decay with a power law.
The exponents  are (weakly) universal in a
range of values of disorder where BH avalanches have a power-law
distributions.
In addition, we find that the distribution of trapping times of
domain wall at a given point in space exhibits scaling behavior.
We show how the long-range noise correlations and fractal properties
of the trapping time distribution are related to the  observed scaling
behavior of BH avalanches.

The present study is motivated by recently observed pattern
formation \cite{Supriya+Barma} and activity correlations along a
driven interface \cite{Sneppen,SOD,Mayaetal} in $1+1$ dimensions
stacked by random defects.
In contrast to these models, here we have a 2-dimensional system with
many interacting interfaces, and  two separate time scales (compared to
extremal driving): slow time scale of field updates, and avalanche
propagation time scale between two field updates. Despite of these
differences, in both cases the
intermittent avalanche-like dynamics occurs, which we believe is
essential for the observed scaling behavior.

\section{Model and simulations}

We consider a simple model with disorder represented by local random fields
$h_x$, which appears from an original disorder via coarse-graining
\begin{equation}
  {\cal{H}} = -\sum_{<x,x^\prime>}J_{x,x^\prime}S_xS_{x^\prime} -
\sum_x (h_x+H)S_x \ ,
\label{Hamiltonian}
\end{equation}
where $x\equiv (x_\|,x_\bot)$ and $J_{x,x^\prime}=1$ is a constant
interaction between nearest-neighbor spins $S_x=\pm 1$.
A Gaussian distribution of $h_x$ is
assumed with zero mean and width $f$.  (Other types of disorder have been
also considered \cite{BT-Calcutta,Vives,BT}). A domain wall of
reversed spins is created  along $<11>$ direction \cite{Uli} on the
square lattice rotated by $\pi /4$. This model is motivated by the
extended domain walls in stress-annealed samples, as discussed  in
Ref.\ \cite{TN}.
Periodic boundaries are applied in the direction of wall and an
open boundary at the opposite side of the wall. The system is driven
{\it globally}---updated value of the external field is applied
to all spins in the system. The dynamics consists of  spin flips when
the local field exceeds zero (see below).

As discussed in Ref.\ \cite{TN}, the fact that motion of the
$\langle 11 \rangle$ domain wall
has no energy threshold at vanishing disorder  has several
advantages. In particular, the wall depinning occurs along
a line $H_c(f) \sim \kappa f$ in the $(f,H)$ plane, where $\kappa $
is a constant (see Fig.\ 2 in Ref.\ \cite{TN}).
 Therefore for the domain wall along $\langle 11 \rangle $ direction
an {\it infinitesimally small} field $H$ is
sufficient to move and depin the domain wall for small disorder,
$H_c(f) \to 0$ for $f\to 0$.
This important property of
the model allows us to use much smaller lattice sizes compared to
earlier studies \cite{JR,cornell}, where a large system has to be used in
order to find a random field large enough to surmount the energy barrier
$2J$ in the case of wall along $\langle 10\rangle $ direction,
or the nucleation energy $4J$ in the case of uniform ground states.
On the other hand, by
increasing disorder the distance between pinning centers decreases,
and thus a critical disorder $f^\star$ exists at which depinning is no
longer possible. Instead, it becomes energetically easier to nucleate new
domains in the bulk. The exact value of $f^\star$
depends on the type of distribution \cite{TN,SD}, and is still not known
exactly.
In Ref.\ \cite{TN} it was found using a finite-size scaling analysis of
the averaged interface velocity and avalanche distributions
that for the present model $f^\star= 0.62$ within numerical error bars.

The scaling properties of BH avalanches in low disorder regime were
discussed in Ref.\ \cite{TN}. Here we are interested in the high disorder
region, where the $\langle 11 \rangle $ interface remains pinned at
all fields. Nevertheless, the initial state with the $\langle 11 \rangle $
 interface ensures that the ratio of the characteristic lengths
 $\xi _1/L \ll 1$ holds for the applied disorder values (thus the
avalanche size cut-off $s_0 \ll L^2$). (The characteristic length $\xi _1$
represents distance between strong pinning centers \cite{JR} for given
strength  of disorder.) On the contrary, when $\xi _1/L
\sim 1$, the $\langle 11 \rangle $ interface moves, meaning that the
selected system size $L$ is small for given strength of pinning, and thus
system is not in the high disorder regime.
Thus, having the condition  $\xi _1/L \ll 1$ satisfied for each applied
value of disorder, we can concentrate on the effects
of disorder fluctuations by applying  many configurations
of random fields at fixed  $f$ and $L$ values. We use up to $L=768$
and up to $10^3$ disorder configurations (i.e., up to $36\times 10^6$ spins).

Since we are interested in the properties of the triggering noise,
which are related to distribution of disorder and the actual dynamics
of the system, we avoid any ``short-cuts'' which can  speed the
algorithm \cite{algorithms}.
We apply the natural slow algorithm, which consists of the
following steps: Random fields are stored at each site of an $L\times L$
lattice; The system is searched for the minimum local field
$h_{min}={\rm min}(h_x^{loc})$,
where $h_x^{loc}\equiv \sum _{x^\prime}J_{x,x^\prime}S_{x^\prime} +h_x +H
\equiv h_{ir,x} +H$,
and then the driving field is set to exactly $h_{min}+\eta $ (we use
$\eta =10^{-10}$), which thus triggers an avalanche at that site;
A list is made consisting of the sites which may flip at the next time
step (neighbor sites to  flipped spins);  The spins on the list are
examined for flip and a new list is made; The process is continued until
no more spins can flip, then next minimum local field is
found.
It should be stressed that all spins in a single list (spin shell) are
updated in parallel, i.e., in a single time step, in analogy to parallel
update in cellular automata models. In this way we have well defined
time scale of avalanche evolution (internal time scale) as the number of
steps that the updating procedure goes before the avalanche stops.
On the other hand, the  slow (external) time scale is set by the number
of driving field updates.

It was shown in Ref.\ \cite{TN} that for low disorder $f<f^\star =0.62$ the
extended domain wall moves through the system and
depinns at a critical field $H_c(f)$, however, above the critical disorder
$f^\star $ the built-in domain wall remains pinned and many new
domains of reversed spins are nucleated inside the system. In this region,
corresponding to high disorder (or low tensile stress), a typical
 structure of clusters which occurs in the linear part of the hysteresis is
shown in Fig.\ 1. The theoretical value $f^\star $ can, in principle, be
related to a critical value of tensile stress
 $\sigma ^\star \sim {f{^\star}}^{\mu}$,  below which the domain
structure changes,
 as also  observed in the experiment in Ref.\ \cite{Brasil}.  More precise
experimental data would be necessary in order to determine the exponent $\mu$.
A rough estimate is that $\mu $ is given by the correlation
length exponent at the transition $f^\star$, $\mu \approx \nu \sim 2.3$
\cite{TN}.

\section{Fractal properties of triggering noise and domain-wall trapping}

 Time series of the magnetization jumps corresponding to the
individual avalanches are known as BH noise. In the high disorder region
either a new domain is nucleated  and grown,
or already existing domain extended, a BH pulse is associated with
motion of a domain wall from one position to a new one. The area between two
consecutive domain wall positions corresponds to the size of BH avalanche
(measured as the area covered by a single BH pulse).
In principle, a domain growth process is not ``linear''
but fractal, leading to the dynamic exponent $z=1.23$, and fractal
 dimension of avalanche $D=1.88$ \cite{TN}.
As usual, the dynamic exponent $z$ and the fractal dimension $D$ are
defined via the scaling of characteristic duration  and size of avalanches
with the change of length scale, $\langle t\rangle _L\sim L^z$, and
$\langle s\rangle _L\sim L^D$, respectively.
Scaling properties are studied in detail (see set of scaling exponents in
Ref.\ \cite{TN}). In particular, in the critical region above $f^\star$
the distribution of size of avalanches $D(s)\sim s^{-\tau _s}$ and duration
$P(t) \sim t^{-\tau _t}$ scale with the exponents
$\tau _s=1.30$ and $\tau _t=1.47$, respectively.
The avalanche distributions integrated over
hysteresis loop for $f>f^\star$ can be scaled according to the scaling
form \cite{TN}
\begin{equation}
P(t,f,L)=t^{-\tau _t}{\cal{P}}(t(\delta f)^{z\nu }, tL^{-z}) \ ,
\label{crit_scal}
\end{equation}
where $\delta f \equiv (f/f^\star -1)$ and $(\delta f)^{-\nu }$ measures
the correlation length due to the critical point at $f^\star$. For instance,
when the condition $L\gg
(\delta f)^{-\nu }$ is satisfied, the avalanche cut-offs can be scaled with
disorder strength ({\it in the entire region of power-law behavior}),
leading to $f^\star =0.62$ and  $\nu=2.3 $ within statistical error bars
(see \cite{TN} for details).
It should be stressed that in this region of disorder the spatial
extension of an avalanche is determined not only by
strength of pinning  but also by the  blocking by previous clusters
(see Fig.\ 1). The blocking effects dynamically alter the strength of
pinning and  thus influence the scaling properties. This
illustrates a complex interplay between the disorder and dynamics,
in contrast to, for instance, the equilibrium clusters in the
random-field Ising model.
In order to get an insight into these complex phenomena, we study next the
distributions of wall trapping times and distances between the  points
with weakest pinning, from which an avalanche is released. We find that both
of these distributions exhibit a scaling behavior.

The distribution of distances between initial points of
the {\it successive} avalanches is
shown in Fig.\ 2, where we distinguish between distances measured
in the parallel ($x_\|$) and transverse($_\bot$)  direction relative to the
direction of the wall. In fact, due to anisotropy of domain wall motion
in this case, these distributions show different scaling
exponents $\tau _\| =1.04 \pm 0.03$ and $\tau _\bot =0.60\pm 0.03$
according to
\begin{equation}
G(x_\|,x_\bot ) \sim x_\|^{-\tau _\|}{\cal{G}}(x_\bot /x_\|^{\zeta _G}) \ ,
\label{GG}
\end{equation}
where $\zeta _G = \tau _\| /\tau_\bot $. Notice that for small distances
correlations in both directions are almost equivalent. However, a crossover
to a distinct power-law behavior for $G(x_\bot)$ occurs at a distance
($x\approx 30$ in Fig.\ 2) which is presumably related to the characteristic
size of avalanche at given disorder. The open boundary in the direction
opposite to wall leads to the cut-off at large $x_\bot$. It should be
stressed that spatial correlations are sensitive to distribution
of disorder. Therefore, here we used a large number of samples in order to
minimize scatter of the data due to disorder fluctuations and to clearly
distinguish between correlations in parallel and perpendicular directions.

In the inset to Fig.\ 2 we show the Fourier spectrum of the threshold
field fluctuations $\Delta H(t)$ measured on the external time scale.
It exhibits two correlated regions
corresponding to   earlier and later times, respectively.
Steeper slope in the inset to Fig.\ 2 vary from 0.6-1.03, and flatter
slope from 0.25 to 0.36, depending on $f$.
The long-range correlations exhibited in Fig.\ 1 show that the system
evolves in such way that its next relaxation event  depends on the
history of the present state of the system.
It is interesting to note that a similar time series of the
fluctuation of  magnetization
$\Delta M(t) \equiv M(t+1)-M(t)$ represents the Barkhausen noise
itself (looked at the external time scale). An example of the BH
noise signal is shown in Fig.\ 3. Note that on the external time scale
duration of each elementary signal is equal to one, whereas the
height of each elementary signal represents corresponding avalanche size.
Sizes of the successive avalanches  are only weakly  correlated in time.
However, a time {\it derivative} of the signal, namely  $a(t) \equiv
d[\Delta M(t)]/dt$, representing acceleration of a domain wall
at each field update, shows certain correlation properties. In the inset to
Fig.\ 3 we show the Fourier spectrum of the numerical derivative of the
the signal, which is shown in main Fig.\ 3. Average slope of the curve
is close to one.

It should be noted that the average number of avalanches occurring  at
fixed strength of disorder increases with system size as $\langle n_a\rangle
\sim L^2/\langle s\rangle \sim L^{[2-D(2-\tau _s)]}\phi (L(\delta f)^\nu)$,
where $\phi (L(\delta f)^\nu)$ is unknown scaling function. This implies that
in the theoretical case of infinitely slow driving the average
size of the field jump $\langle \Delta H(t)\rangle $ decreases with $L$ as
$\langle \Delta H(t)\rangle  \sim H_{sat}(L)/\langle n_a
\rangle \sim L^{-0.68}H_{sat}(L)/\phi (L(\delta f)^\nu)$, where $H_{sat}$ is
the saturation field. In real experimets
the size of field jumps are restricted by the driving frequency, however,
number of avalanches detected varies with the size of pick-up coil.
This problem requires  more detailed theoretical and experimental
investigation \cite{Djole}.


Further understanding of the role of disorder in the dynamics  can
be achieved by considering the time intervals that a domain wall resides
at a given point in space (trapping time $T_{trap}$).
 We calculate the distribution of trapping times $T_{trap}$
of a domain wall at a given site, which is determined as the time interval
since the domain wall is pinned at a site $(x_\|,x_\bot )$
until it eventually moves away from that site. In this way, $T_{trap}$
measures the time interval between two successive activities at that
site.
Notice  that $T_{trap}$ in this case is somewhat different from  so called
first-return time in 1-dimensional interface \cite{Sneppen}.
Here we have 2-dimensional system with many interacting interfaces.
Another similar example is trapping of grains in rice-pile models \cite{Al}.
A reasonable time scale to measure  $T_{trap}$ is the number of field updates
(external time scale), since in the zero-temperature dynamics during an
avalanche a spin at a given site is either fixed or flips only once.

The trapping-time distribution
is shown in the inset to Fig.\ 4 for various values of disorder $f >f^\star$.
Between the lower cut-off $T_0$ (below which all trapping times are equally
probable), and an upper cut-off due to lattice size $L$,
the trapping time distribution shows a short region with correlations.
In particular, for large trapping times we can determine the slope  as
$\tau _{trap}=2.3 \pm 0.1$.
One can understand that  a  metastable configuration of domain walls
corresponds to the system residing in a ``pocket'' of the  fractal free
energy landscape. A domain wall may move from a site $(x_\|,x_\bot)$ as
a part of an avalanche which started in the neighborhood of that point,
corresponding to a local reorganization of the landscape near a shallow
minimum (for a given value of the external field). When the system resides
in a deeper minimum, however, it waits for larger driving  fields
({\it i.e.}, trapping time increases) or for a more global
reconstruction of the landscape, which  occurs with smaller probability
 compared to the one discussed above. This leads to the large slope
of the distribution $P(T_{trap})\sim T_{trap}^{-\tau _{trap}}$
at large $T_{trap}$.
It is interesting to note that a similar slope
for large trapping times was found in the case of rice-pile model \cite{Al},
where trapping of grains are considered, and it was argued that the slope
is related to the roughness  exponent of the rice-pile surface
(see detailed analysis in Ref.\ \cite{Al}).
Analogies between an interface motion and rice-pile model have been
established in the literature \cite{Maya}. Considering the
avalanche exponents the analogy also applies
 for the BH noise with an extended domain wall in the limit of low disorder
\cite{TN}.
Notice that, in contrast to
ricepile model, the range of power-law behavior in BH noise depends
not only on the system size, but also on the strength of disorder, which
restricts spatial extension  of avalanches to $s\le s_{max}\ll L^2$ and
their durations to $T\le T_{max}$.
Therefore, in this region of disorder we have rather small range of the
power-law behavior. However, the scale-free
behavior of the distribution of trapping times can be demonstrated via
finite size scaling analysis when different system sizes are used. We find
that the  following scaling form

\begin{equation}
P(T_{trap},L) = L^{\alpha }{\cal{P}}(T_{trap}L^{-z_T}) \ ,
 \label{fss}
\end{equation}
applies  with the exponents which are weakly dependent on disorder
$\alpha =-0.4\pm 0.05$ and $z_T=1.66 \pm 0.05$ (see Fig.\ 4).
By increasing the system size $L$ the trapping times increase,
leading to larger cut-offs of the distribution. We notice that the
lower cut-offs $T_0$ also increase, leaving a limited correlation range.
However, the cut-offs scale nicely with $L$, as shown in Fig.\ 4.
In the inset to Fig.\ 4 we show how the trapping times distribution
varies with disorder.
 By increasing disorder the cut-off $T_0$ moves towards larger values
and the correlation region shrinks, corresponding to a lesser
correlations in the dynamics, which  also manifests in the decrease
of the characteristic avalanche size. In the limit
of infinitely strong  disorder the dynamics becomes completely random,
consisting of individual spin flips which align along  a local
random field.

It is interesting to note that unlike the avalanche distributions in
Eq.\ (\ref{crit_scal}), the distribution of trapping times in Eq.\
(\ref{fss}) as well as $G(x)$ (\ref{GG}) do not show explicit disorder
dependence apart from a weak dependence in the exponents. In fact, the
disorder effects in this case are included in the gradient of driving
force $\Delta F$ (i.e., the increments
of the external field $\Delta H$, which are adjusted to the
minimum local fields). For example, for the distribution of trapping times
we have in general
\begin{equation}
P(T_{trap},L,\Delta F) =
L^\alpha {\cal{P}}(T_{trap}(\Delta F)^{z_T/\lambda _F},T_{trap}L^{-z_T}) ,
\label{general}
\end{equation}
where the increase in the driving force $\Delta F \equiv H(t+T_{trap})-H(t)$
contains all the contributions due to interaction and random pinning
which occurred during the time interval $(t,t+T_{trap})$, i.e.,
$\Delta F = \sum _t^{t+T_{trap}}h_{ir,x}$,  which steam from the different
sites in the system.
According to the above results in Fig.\ 2, sum of these contributions has no
characteristic scale.
Therefore we may conclude that $(\Delta F)^{-1/\lambda _F} \to \infty $.
Hence the first argument in the right hand side of Eq.\ (\ref{general})
can be neglected
compared to $T_{trap}/L^{z_{trap}}$, leading to the scaling form (\ref{fss}),
which is in agreement with the scaling plot in Fig.\ 4.

Another interesting observation regards the
 dynamic exponent $z_T$, which scales the tail of the trapping time
distribution with the system size $L$. It may be related to the
scaling of the length of the
optimal path \cite{opt-path} in strongly disordered medium, $z_T=D_{OP}$.
 In fact, the optimal path between two points can  be constructed from
the {\it most persistent sections} of the
domain walls in the dynamics of BH avalanches. The length of the
optimal path scales with the linear distance between the end points
with the exponent $D_{OP}= \tau _{OP} = 1.66$
\cite{BT-RG,opt-path}.

\section{Transport equation with noise correlations}

The long-range noise correlations are shown to be relevant for the
scaling properties of the interface motion\cite{IF-noise}. In the
literature the  noise correlations
in the interface depinning are viewed as originating from
another (external) dynamic  processes \cite{IF-noise}.
In the case of Barkhausen avalanches we see that the noise correlations
appear as an intrinsic property of the dynamics when the system is driven
infinitely slowly.
 The active section of domain wall can be  represented
by a surface $h(x,\tau )$  which is pinned by quenched defects.
The transport at the site $(x,\tau )$ is described by the equation
\begin{equation}
dh/d\tau = \nu _\| \partial _\|^2 h + \nu _\bot \partial _\bot^2 h
+ \eta (x,t,h) \ .
\label{transport}
\end{equation}
Here we distinguish total elapsed  time  $\tau $, which is defined
as the  sum of evolution times of all individual avalanches, on one hand,
and $t$, which represents time scale of field updates, on the other.
The quenched noise $\eta (x,t,h)$  is
generated by varying the driving field in the steps which are adjusted
to the minimum local strength of pinning, as discussed above.
We expect that a dominate $h$-dependence of the noise $\eta $ can be
expressed as $\eta \sim p(x)\partial h $, where $p(x)$ represents the
(anisotropic) local velocity of domain wall per field rate. In addition,
time variation of the noise are related to the updates of the external
field, and thus $\eta $ varies on the external time scale only. Therefore,
we can write
\begin{equation}
\eta (x,t,h) \approx  p(x)(a_0 + \Delta H(t))(\partial h) \; ,
\label{eta}
\end{equation}
where $a_0$ is a constant which
depends on initial configuration and the hysteresis loop properties.
The local interface velocity per field rate, $p(x)$, is a random variable
 which is determined by the spatial distribution and strength of pinning
associated with  a given value of the driving field $H$, i.e.,
history of the system. Therefore $p(x)$ is governed
by a probability distribution, which can be deduced from the properties
of the probability that an avalanche starts at distance $x$ from
the preceding avalanche, i.e.,
\begin{equation}
<p(x)p(x^\prime)> = \gamma G(x-x^\prime) \; .
\label{pp}
\end{equation}
It is assumed that the same type of correlations apply for the
successive activities during the evolution of an avalanche, while the
external field is constant.
In the previous section we found that the probability $G(x)$ exhibits
long-range correlations in the perpendicular and parallel direction
as $G(x_\|,x_\bot) \sim x_\|^{-1}x_\bot^{-0.60}$. We also notice that
$\Delta H(t)\sim t^{-\theta }$, where $\theta $ is finite in the case of
infinitely slow driving discussed  above, whereas, $\theta =0$ in the
case of finite driving rate in small steps \cite{BT}, where
$\Delta H(t) =const$. Notice that
Eq.\ (\ref{transport}) together with Eqs.\ (\ref{eta}) and (\ref{pp})
leads to an effective nonlinear term of the form
$\gamma G(x_\|,x_\bot)(a_0+t^{-\theta })^2(\partial h)^2$, which
differs from the Kardar-Parisi-Zhang
 nonlinearity \cite{KPZ} by the power-law correlations in the coefficient.
As discussed above, the origin of these correlations
lies in the the dynamically varying pinning and blocking effects
when a multidomain structure is slowly driven through the hysteresis loop.

The relevance of the noise correlations for the universality class of the
interface depinning has been discussed  in
the literature using dynamic renormalization group (RG)
\cite{IF-noise}. Another example where correlations of random noise play
an important role is represented by the scaling properties of river
networks \cite{BT-RG,RN}. In that case the correlations of the type
$G\sim x_\|^{-2+\delta }x_\bot ^{1-\zeta}$ were assumed, where $\delta$ is
an expansion parameter and the anisotropy exponent $\zeta $ has to be
determined self-consistently at a fixed point of the RG \cite{BT-RG}.
In the result, the scaling exponents do depend on the range of
correlations $\delta$.
Similarly,  we can expect that the long-range correlations
of the form given in Eqs.\ (\ref{eta}) and (\ref{pp}) can be related to the
universal scaling exponents of the BH noise.
The exponents
 $z$, $\zeta _G$, which are defined above,  and the roughness exponent
$\chi$, which governs behavior of the dynamic variable $h$ with the change
of scale, can in principle be obtained by the dynamic RG  applied to
Eq.\ (\ref{transport}) with noise properties in Eqs.\ (\ref{eta}-\ref{pp}).
Then the avalanche exponents $\tau _t$ and $\tau _s$ can be deduced using
scaling relations in the critical region. The RG analysis of the transport
equation (\ref{transport}) with noise properties specified by
Eqs.\ (\ref{eta}-\ref{pp}) requires additional work and is left out of
the scope of this paper. Here we discuss only the scaling relations
between the exponents at RG fixed point, e.g., the dynamic exponent $z$,
 and the avalanche exponents.

It should be stressed that such scaling relations are not universal and that
they depend on the nature of the dynamic process
(see for instance \cite{BT-RG,RN} for the case of river networks).
We argue that the scaling relation derived below are valid for
BH avalanches in the high-disorder (or multidomain) region.
The rational behind these scaling relations is found in the directed
nature of the avalanche propagation. The evolution of an avalanche
(see Fig.\ 1) can be visualized as growth of a directed percolation
(DP) cluster projected \cite{comment} to $1+1$ dimensions, with extra
dimension representing time axis.
The fractal dimension of the equivalent DP cluster measured with respect to
the time axis is then $D_\| \equiv D/z$, where $D$ and $z$ are fractal
dimension of BH avalanche and dynamic exponent, respectively,
as defined above. Then $D_\|=1+\zeta _{DP} -\delta _{DP}$, where
$\zeta _{DP}=z/2$ is the anisotropy exponent measured with respect to
time axis, and $\delta _{DP}$ is the survival probability
exponent of the equivalent  DP clusters.
Notice also that for the directed dynamic processes $\tau _t =D_\|$ and
thus $\tau _s=2-1/\tau _t$ \cite{TD}, which completes  the set of
avalanche exponents. In addition, the roughness exponent $\chi $
can be related to the trapping time distribution as $\tau _{trap} =2+\chi $,
as noticed in Refs.\ \cite{Al,Maya}. For instance, using the numerical
values for $D$ and $z$ from Ref.\ \cite{TN}, the following values of
the avalanche
exponents are predicted by the above scaling relations: $\tau _t=1.52$,
$\tau _s=1.34$, compared with $1.47$ and  $1.30$, and $\chi =0.23$,
 obtained by direct numerical simulations.

\section{Conclusions and discussions}

We have demonstrated numerically the existence of long-range correlations
in triggering noise which is intrinsic to the domain wall dynamics in
slowly driven disordered ferromagnets. Although the distribution of
disorder is initially uncorrelated, the spatio-temporal correlations
develop in time. The appearance of these  correlations can  be
attributed to the applied global driving in which always
{\it next weakest} pinning force in the system is selected, on one hand,
and to  a finite extension of avalanches occurring between two
consecutive field updates, on the other.
Hence, attempts to reduce the domain wall dynamics to the Glauber
spin flips in the presence of local random fields appears to be
inadequate for the range of disorder where the cooperative avalanche
dynamics occurs.
We suggest an alternative transport equation which incorporates the
observed noise correlations.
 The long-range correlations of triggering noise can be related to the
universal scaling behavior of Barkhausen avalanches and to the transport
properties {\it via} the fractal distribution of the domain wall
trapping times.
We believe that an  analysis of the transport equation by the
dynamic RG will contribute to understanding  of the universality
classes of Barkhausen avalanches by the infinitely slow driving.
By varying the driving conditions, however, these correlations are
 changed. This might be the origin of different scaling behavior
of BH avalanches at finite driving rates, as observed in experiments
\cite{Durin-stress}.
Our results also suggest that technical procedures which
speed the numeric algorithms in driven disordered systems
should be taken with care. In particular, an algorithm which alters
the properties of triggering noise may lead to a different scaling
behavior, which is unrelated with the original problem.

\acknowledgments
This work was  supported by the Ministry
of Science and Technology of the Republic of Slovenia.
I am grateful to Deepak Dhar for many helpful comments and suggestions.


\narrowtext

\narrowtext
\begin{figure}
\epsfxsize=80mm\epsffile[21 255 589 537]{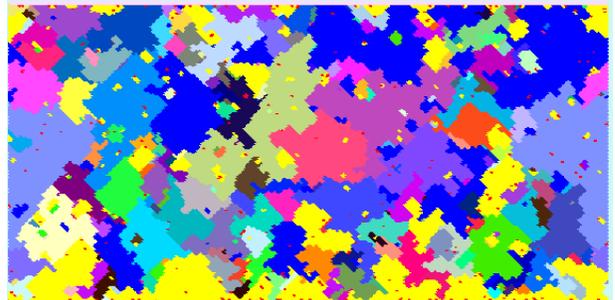}
\caption{Extended domain wall along the bottom line (shown in red), and
 unflipped spins (full blue color) and
cluster of flipped  spins developed  at different stages of evolution
(shown using a continuous color map [23]).
Only 253 recent clusters are shown, with all older clusters shown in
yellow color. $L=200$ and $f$=1.1 .
}
\label{fig1}
\end{figure}
\begin{figure}
\epsfxsize=82mm\epsffile[47 68 522 581]{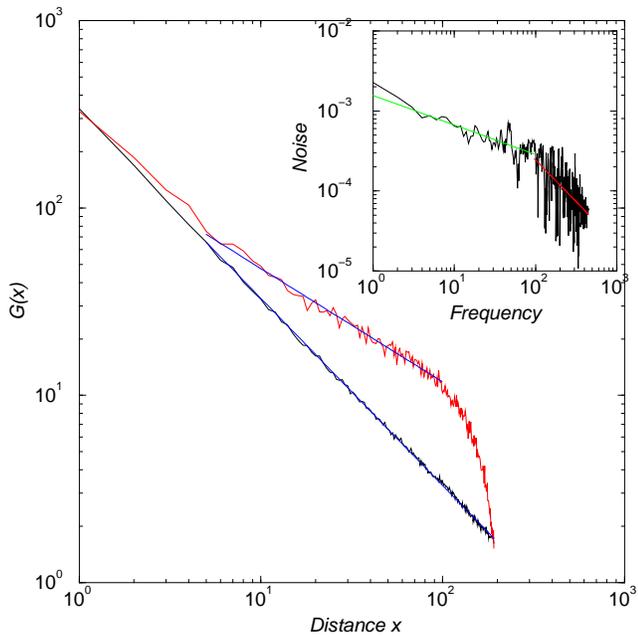}
\caption{The probability that the subsequent avalanche starts at the
distance $x_\|$,$x_\bot $ (bottom, top) from the point of the preceding
avalanche for  $L=192$ and $f=1.0$. The data are averaged over 1000 samples.
Inset: Fourier spectrum of the driving field fluctuations $\Delta H(t)$
taken in {\it one} ascending branch of hysteresis for $f=1.0$ and $L=192$,
vs. undimensional frequency. }
\label{fig2}
\end{figure}

\begin{figure}
\epsfxsize=82mm\epsffile[54 71 565 393]{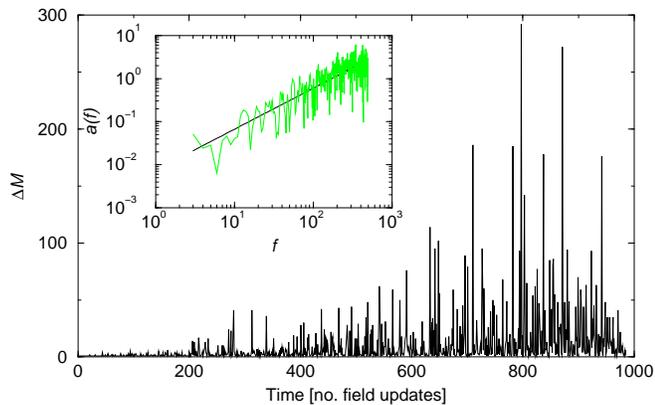}
\caption{Sequence of Barkhausen pusles $\Delta M$ (measured in
number of spins) recorded on the time scale of field updates.
Inset: Fourier spectrum of the numerical derivative of the sequence
in the main figure plotted against undimensional frequency. Fitted
slope is $0.95\pm 0.03$.}
\label{fig3}
\end{figure}

\begin{figure}
\epsfxsize=82mm\epsffile[47 68 522 581]{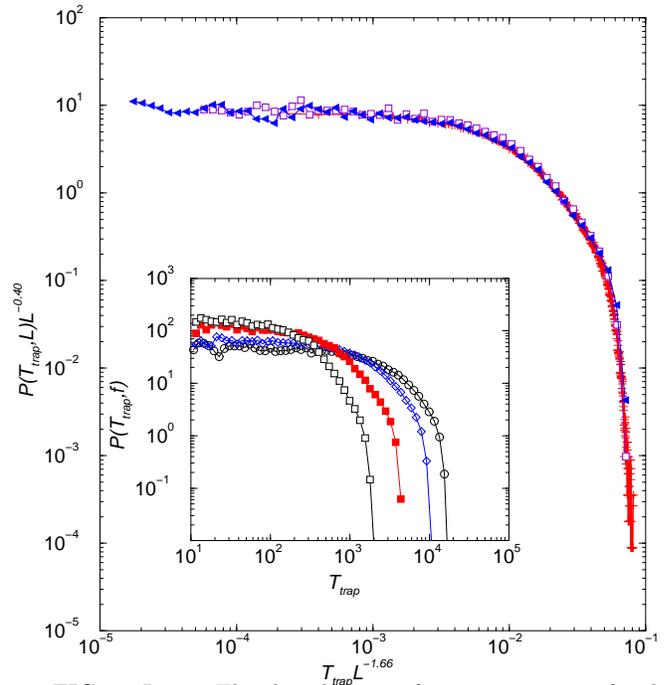}
\caption{Inset: The distribution of trapping times of a domain wall vs.
trapping time $T_{trap}$ measured by number of field updates.
The distributions for various values of disorder $f$=0.9, 1.0, 1.1, and 1.2
(left to right) are obtained at lattice $768\times 768$ averaged over
10 samples. The data are logarithmically binned. Main: Finite size scaling
plot of the trapping time distribution for $f$=1.0 and the linear lattice
size $L$=192, 384, and 768. The data for $L=192$ are averaged over 1000
samples, the rest of the data over 10 samples, and are logarithmically binned.}
\label{fig4}
\end{figure}


\end{multicols}


\begin{references}
\bibitem{comment} Such projections are manifestly used in the problem of
river networks \cite{BT-RG,RN}.
\end{references}
\end{document}